\documentstyle[11pt,paspconf]{article}

\input psfig

\def\approxgt{\,\raise2pt \hbox{$>$}\kern-8pt\lower2.pt\hbox{$\sim$}\,}
\def\approxlt{\,\raise2pt \hbox{$<$}\kern-8pt\lower2.pt\hbox{$\sim$}\,}

\def\ni{\noindent}
\def\th{\thinspace}
\def\sss#1{{{\scriptscriptstyle #1}}}
\def\ah{{\scriptscriptstyle 1/2}}
\def\Teff{{$T_{ef\!f} $}}
\def\Mo{{M$_\odot $}}
\def\Lo{{L$_\odot $}}
\def\eg{{e.g.~}}
\def\cf{{cf.~}}
\def\ie{{i.e.~}}
\def\etal{{et al.~}}

\begin{document}

\title{Turbulent Convection in Pulsating Stars}
\author{J. Robert Buchler, Philip Yecko, Zolt\'an Koll\'ath, Marie-Jo Goupil}
\affil{Physics Department, University of Florida,
    Gainesville, FL 32611}

\begin{abstract}
We review recent results of stellar pulsation modelling that show 
that even very simple one-dimensional models for
time dependent turbulent energy diffusion and convection provide a substantial 
improvement over purely radiative models.

\ \ 

{\bf
\ni Workshop on Stellar Structure: \th
Theory and Tests of Convective Energy Transport,\th\th 
Granada, SPAIN, 1998

\ni (to appear in ASP Conference Series)}

\end{abstract}


\section{Introduction}

Cepheid and RR Lyrae modelling has a long history going back to the early
1960's (recently reviewed in Gautschy \& Saio 1995).  Right from the beginning
it was quite clear that convection had to be present in the pulsating
envelopes.  Furthermore, modelling showed that convection was necessary to
provide a red edge to the instability strip, \ie to stabilize the stars at
lower temperatures.  However convection was deemed to have a minor effect on
the shape of the light curves and radial velocity curves.  And, indeed, purely
radiative models gave good agreement with the observations of the Galactic
Cepheids (\eg Moskalik \etal 1992), although a few problems of varying degree
of severity persisted (\cf Buchler 1998), such as the inability of radiative
codes to model beat pulsations in either Cepheids or RR Lyrae.  The light curves
of the so-called Beat Cepheids or RR Lyrae indicate that these stars pulsate
with two basic frequencies, and with constant power in these frequencies.  In
addition, radiative codes give pulsation amplitudes that are much too large
when compared to the observations.  Furthermore the amplitudes depend on the
fineness of the numerical mesh.  The amplitudes as well as the stability of the
limit cycles also depend on the values chosen for the pseudo-viscosity.

In the last few years a wealth of data on variable stars in the Magellanic
Clouds (MC) has been obtained as a by-product of the EROS and MACHO
microlensing projects.  Because these galaxies have a metal content that is
only one quarter to one half that of our Galaxy, our observational data base
has therefore been substantially broadened.  Calculations with radiative codes
show rather clearly that {\sl purely radiative models} are incapable of
agreement with observations (\eg Buchler 1998).

The fact that resonances among the vibrational modes give rise to observable
effects (\eg Buchler 1993) can be exploited to put constraints on the
pulsational models and on the mass--luminosity relations.  The best known of
these resonances occurs in the fundamental Cepheids ($P_0/P_2$=2) in the
vicinity of a period $P_0$=10 days and is at the origin of the well known
Hertzsprung progression of the bump Cepheids.  MC observations (Beaulieu \etal
1995, Beaulieu \& Sasselov 1997, Welch \etal 1997) show that the resonance may
be slightly shifted to half a day or a day higher in period.  Structure also
appears in the Fourier decomposition coefficients of the first overtone Cepheid
light curves, and is most likely due to a resonance $P_1/P_4$=2 with the fourth
overtone as first pointed out by Antonello \etal (1980).  Again MC observations
indicate that the resonance center occurs approximately at the same period.
When used to constrain purely radiative models (Buchler, Koll\'ath, Beaulieu \&
Goupil 1996) one obtains stellar masses that are much too small to be in
agreement with stellar evolution calculations.

Improvements to the radiative Lagrangean codes have been made in recent years:
Adaptive mesh techniques have been used to resolve sharp spatial features
such as shocks and ionization fronts.  Instead of treating radiation in an
equilibrium diffusion approximation, the equations of radiation hydrodynamics
have been implemented.  However, all these changes have not substantially
improved the agreement between modelling and observations.  It has become
patently clear that some form of convective transport and of turbulent
dissipation is needed if we want to make progress.

\begin{figure}
\centerline{\psfig{figure=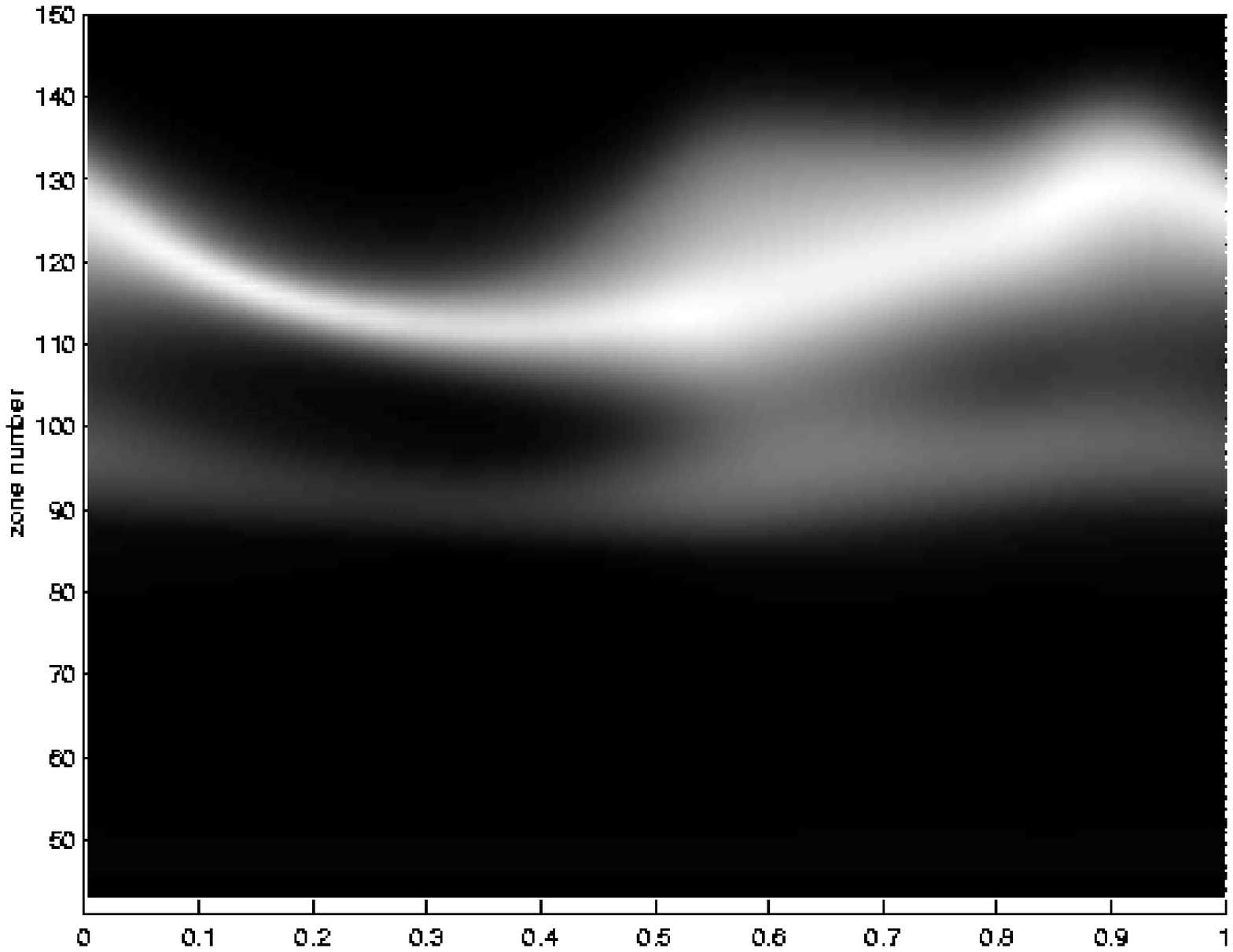,width=6.5cm}
\psfig{figure=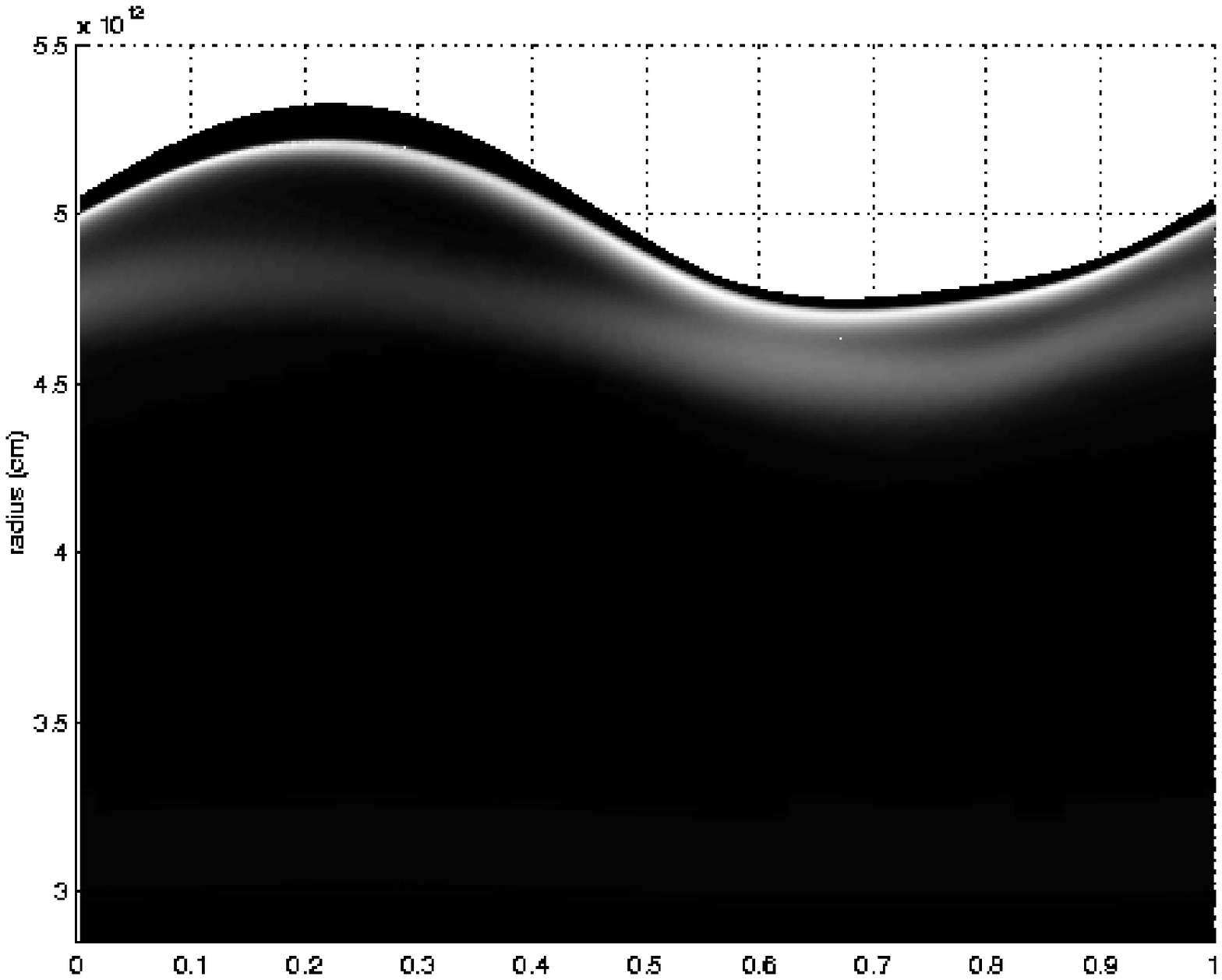,width=6.5cm}}
\caption{\small Time dependence of turbulent energy during the pulsation 
cycle.\th  Left: vs. zone index, right: vs. Lagrangean radius.} \label{figet}
\end{figure}

Turbulence and convection are inherently 3D phenomena.  While a great deal of
progress has been made in 3D simulations, it remains very difficult to model
astrophysically realistic conditions which have very large Rayleigh numbers
Ra$\approx$ 10$^{12}$, and very small Prandtl numbers Pr$\approx$ 10$^{-6}$.
It is of course even more difficult to incorporate them in stellar models (see
however the solar models of Nordlund \& Stein at this meeting).

Large amplitude stellar pulsations increase the difficulties by involving time
dependence. Indeed, the source regions occur in the partial ionization regions
of hydrogen, helium and Fe-group atoms, and these features are neither
Lagrangean (they move through the fluid) nor Eulerian (they move through
space).  Fig.~\ref{figet} shows the behavior of the turbulent energy $e_t$ over
a period in a pulsating Cepheid model with a period of 10.9~days ($M$=6.1\Mo,
$L$=3377\Lo, \Teff=5207K, $X$=0.70, $Z$=0.02).  Similar behavior occurs in RR
Lyrae models, but we concentrate here on Cepheids which are actually more
daunting numerically because of the sharpness of their H ionization front.  On
the right side we display $e_t$ as a function of Lagrangean radius.  The
lightness of the grey reflects the strength of $e_t$.  On the left,
Fig.~\ref{figet} displays $e_t$ as a function of zone index, \ie as attached to
the Lagrangean mass coordinate, \ie in the fluid frame.  The turbulent energy
is largest in the region associated with the combined H and first He ionization
fronts.  The next most important region of turbulent energy is the
He$^{\scriptscriptstyle +}$--He$^{\scriptscriptstyle ++}$.  There can also be
turbulent energy in the Fe group partial ionization regions, at least for
Galactic metallicity, but is comparatively weak and does not show up on the
scale of the figure.  Fig.~\ref{figet} clearly shows how the turbulent energy
tracks the source regions which move through the fluid during the pulsation.
It also shows the importance of time dependence in the convective pulsating
envelope.  Both figures show that the turbulent energy increases during the
pulsational compression phase and that the two turbulent zones briefly merge.

\begin{figure}
\vskip 10pt
\centerline{\psfig{figure=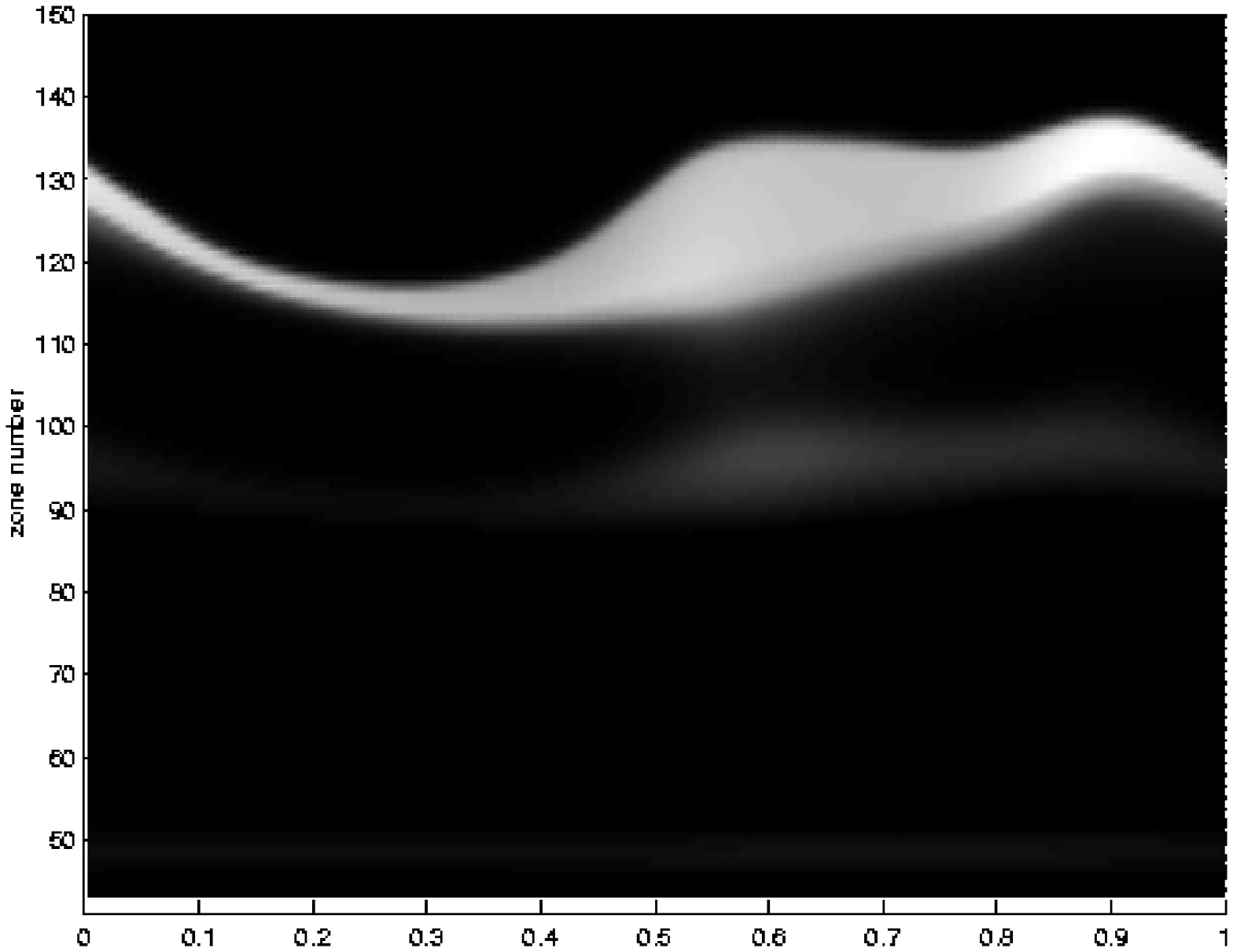,width=6.5cm}
\psfig{figure=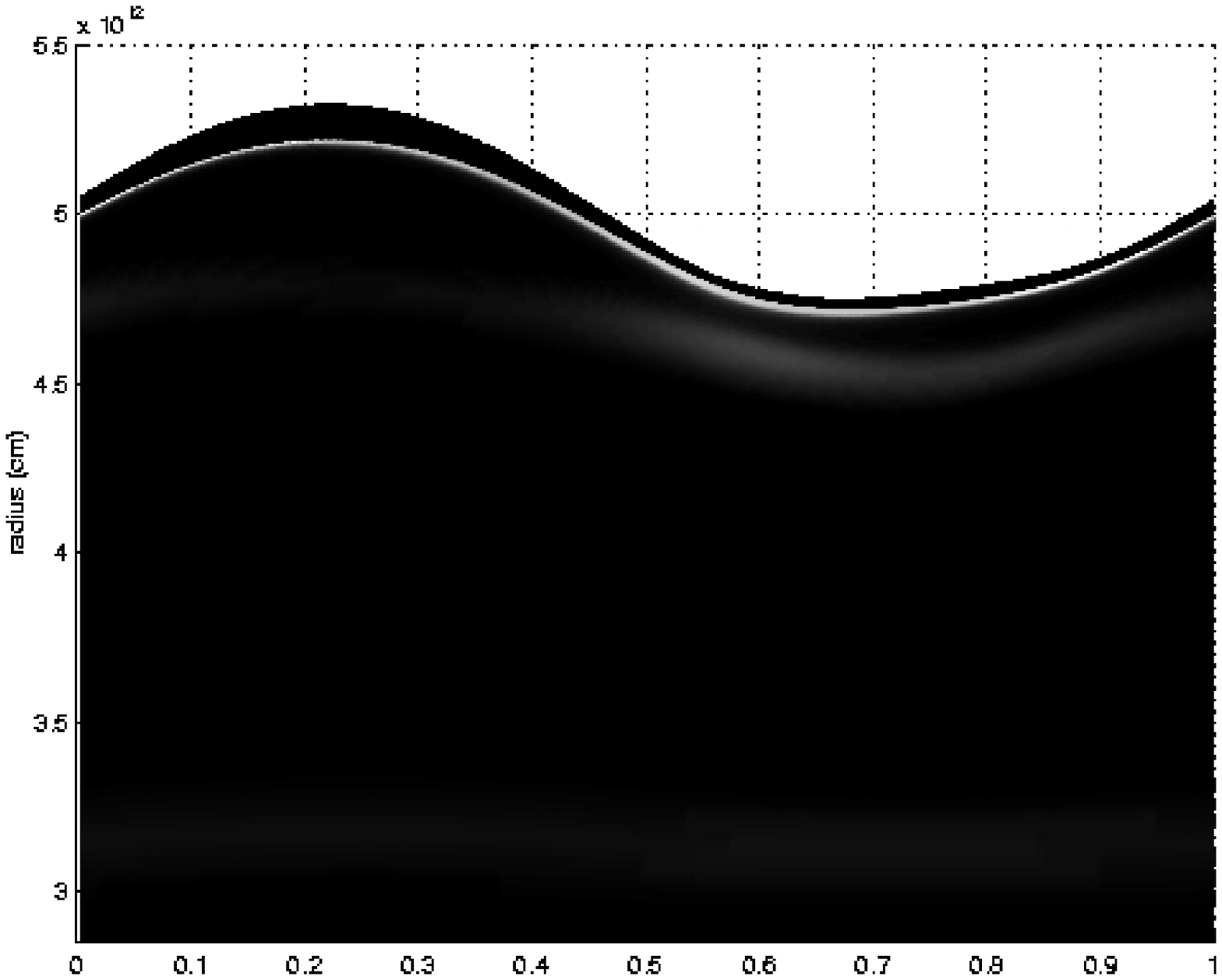,width=6.5cm}}
\caption{\small Convective luminosity; left: vs. zone index, right:
vs. Lagrangean radius.} \label{figlc}
\end{figure}

Fig.~\ref{figlc} shows the temporal behavior of the convective flux in the
frame of the zone index (Lagrangean) and in the stellar frame, respectively,
and can be compared to the turbulent energy in Fig.~\ref{figet}.  The
convective flux exists only in the regions of negative entropy gradient regions
($Y>0$, \cf Eq.~\ref{sk}) and is therefore confined to narrower regions than
the turbulent energy which can diffuse outside these regions.

\hyphenation{Stelling-werf}

 Attempts at including convection in pulsation codes
are actually not new.  Castor (1971) was the first to present a nonlocal time
dependent formulation and numerical application to a pulsating stellar envelope
model.  Later, Stellingwerf simplified Castor's formulation and performed a
number of model calculations (Stellingwerf 1982, Bono \& Stellingwerf 1994).
Similar turbulent convective model equations have been used by Gehmeyr and
Winkler (1992).  Another early computation of linear convective models is that
of Gonczi \& Osaki (1980).

\section{The Turbulent Convection Recipe}

 \begin{equation}
   {du\over dt} = -{1\over\rho}{\partial \over\partial r}
\left(p+p_t+p_\nu\right)
   - {G M_r\over r^{\scriptstyle 2}} \, ,
\label{eqmom}
 \end{equation}
 \begin{equation}
   {d\over dt}\left(e+e_t\right) =
   -{\left(p+p_t+p_\nu\right)\over\rho}\
  {1\over r^{\scriptstyle 2}} {\partial r^{\scriptstyle 2} u\over\partial r}
   -{1\over\rho r^{\scriptstyle 2}} {\partial \over\partial r} \left[ r^{\scriptstyle 2}
\left(F_c+F_t+F_r\right)\right] \, ,
 \label{eqene}
 \end{equation}
 \begin{equation}
  {de_t\over dt} = -{1\over\rho r^{\scriptstyle 2}} 
{\partial \over\partial r}\left( r^{\scriptstyle 2} F_t\right)
  - { e^\ah_t\over\Lambda}
  \alpha_d\th \left(e_t - S_t\right)
   -{\left(p_t+p_\nu\right)\over\rho}\
  {1\over r^{\scriptstyle 2}} {\partial r^{\scriptstyle 2} u\over\partial r} \, ,
\label{eqete}
 \end{equation}
 \begin{equation}
  p_t = \alpha_p \th \rho \th e_t \th ,\quad\quad\quad
  p_\nu = - \alpha_\nu\th \rho \Lambda
  e^\ah_t  {\partial u\over\partial r} \, ,
  \end{equation}
 \begin{equation}
  F_t =-\alpha_t\th \rho \Lambda {2 \over 3} {\partial\th\th e_t^\sss{3/2}
\over\partial r},
 \quad\quad   F_c = \alpha_c\alpha_\Lambda\th \rho e_t^{\ah} \th  c_p T\th Y
 \, ,
 \end{equation}
 \begin{equation}
  S_t = \alpha_s\th \alpha_\Lambda\th ( e_t\th {p\over \rho} \th\beta T\th Y
)^{\ah}, \ \ \ \ \
 \label{eqoldgw}
 Y = \left[ -{H_p \over c_p} {\partial s \over \partial r}\right]_+ \label{sk}
 \, ,
 \end{equation}
 where $p$ is the gas pressure, $\beta$ is the thermal expansion
coefficient, $\Lambda = \alpha_\Lambda\th H_p$, and $H_p = p\th
r^{\scriptstyle 2}/(\rho GM)$ is the pressure scale height, and other 
symbols have their usual meanings.

This scheme gives rise to an unphysical behavior at the boundaries of the
convective regions where $Y\rightarrow 0$.  Because $\delta S_k \propto \delta
Y /\sqrt Y$ the linearization has a pole there that shows up in the growth
rates along sequences of models when the zoning is very fine or the mesh
happens to fall on a point with small $Y$.

\begin{figure}
\centerline{\psfig{figure=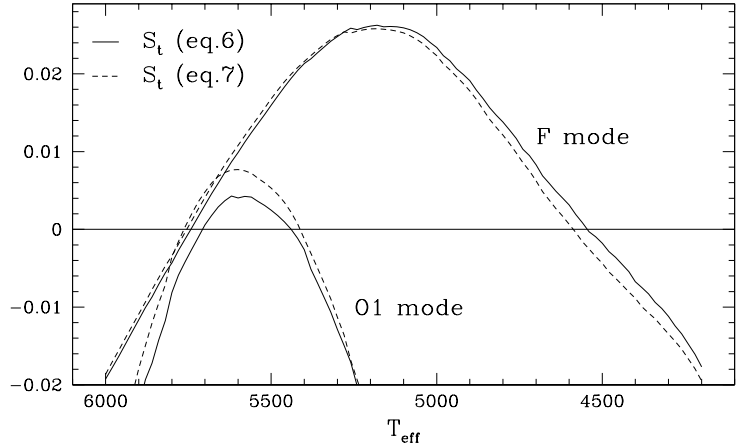,width=10cm}}
\caption{\small Fundamental and first overtone growth rates for a
Cepheid sequence. Solid line: formulation of Eq.~\ref{eqoldgw}, dotted line:
Eq.~\ref{eqgw}.}
\label{figgw}
\end{figure}

This difficulty can be avoided with an alternative model equation for the
source which is more in line with the Gehmeyr--Winkler formulation (1992)
\begin{equation}
 S_t = (\alpha_s\th \alpha_\Lambda)^{\scriptstyle 2} \th\th
{p\over \rho} \th\beta T\th\th Y \, , \label{eqgw}
 \end{equation} 
For comparison, in Fig.~\ref{figgw}, we show the effect that the two
formulations have on the linear growth rates for a sequence of Cepheid models
($M$=6.75\Mo, $L$=4843, $X$=0.70, $Z$=0.02, variable \Teff).  The difference is
seen not to be substantial, although the alternate $S_t$ increases the maximum
period that unstable overtone models can have (\cf Fig.~12 and 13 in YKB).  For
this model we have taken the parameters ($\alpha_d = 1.0, \alpha_c = 2.25,
\alpha_s = 0.75, \alpha_\nu = 1.8, \alpha_t = 0.25, \alpha_p = 0.667, e_0 =
1.e4, \alpha_\Lambda = 0.4$).  These values are used for illustrative purposes.
Except for this one example we have not yet explored the general effects of
using expression (\ref{eqgw}).  We are also still in the process of calibrating
the $\alpha$'s using the best available astrophysical constraints.

\section{Work Integrand}

It is interesting to see how turbulent convection affects the stability of the
pulsational modes.  Turbulent convection actually affects the stability in two
ways, indirectly, by altering the structure of the equilibrium model and,
directly, in the linearization of the equations.  (\cf \eg Yecko, Koll\'ath \&
Buchler 1998, YKB hereafter).

The work done on the pulsation per cycle  is given by
 \begin{equation}
 W = \oint dt \th\th \bigl (\int   dm\th p {dv \over dt}\bigr ), \label{eqwk} 
 \end{equation}
 where the total pressure $p=p_g +p_t +p_\nu$ is composed of the separate
contributions of the gas (and radiation) pressure $p_g$, the turbulent pressure
$p_t$ and the eddy viscous pressure $p_\nu$.  If we denote the linear
eigenvalues by $\sigma = i \omega +\kappa$, for an assumed exp($\sigma t$)
dependence, then the relative growth rate is given by
 \begin{equation}
 \eta =   2{\kappa\over \omega}
    ={2\pi \over \omega^{\scriptstyle 2} I} \th\th 
{\rm Im} \int \delta p\delta v^* \th dm \th ,
 \end{equation}
 \begin{equation}
 I = \int \vert\delta r\vert^{\scriptstyle 2} \th dm \th ,
 \end{equation}
 and the $\delta$ refer to the pressure, specific volume 
and radial displacement parts
of the modal eigenvector, respectively, and $I$ is the moment of inertia of 
the mode.

The quantity $\eta$ represents the energy growth of a mode over one period,
equal to the inverse of the quality factor $Q$ that is commonly associated with
resonant electronic devices.

Here we illustrate with a fundamental Cepheid model (with M=5.2\Mo, L=3293\Lo,
\Teff =5677\th K, X=0.716, Z=0.01) how the work integrands are affected by
convection.  It is of interest to see how the various regions contribute to the
work, as well as how the turbulent convective
quantities affect the stability.

\begin{figure}
\centerline{\noindent\psfig{figure=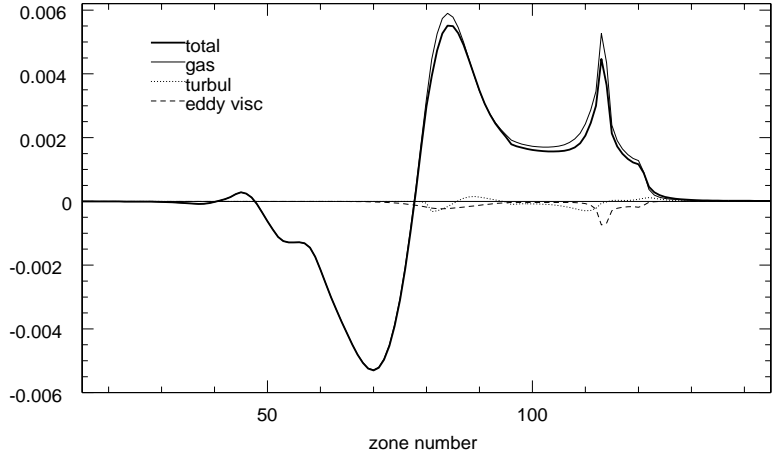,width=10.5cm}}
\caption{\small Linear Work Integrand (surface to the right) showing separate
contributions of the gas, turbulent and eddy viscosity pressures.}
 \label{linwk}
\end{figure}

In Fig.~\ref{linwk} we display the {\sl linear} work integrand (thick solid
line) together with the separate contributions: gas pressure (thin solid line),
$p_t$ (dotted line) and $p_\nu$ (dashed line).  The area under the curve is
slightly positive since the mode is linearly unstable.  As expected, the eddy
pressure is everywhere damping.  The turbulent pressure, on the other hand, can
be both driving or damping depending on its phase with respect to the density
variations.  The sharp peak is associated with the H and first He ionization,
whereas the broad peak is due to the second He ionization and the tiny peak to
Fe.

\begin{figure}
\centerline{\psfig{figure=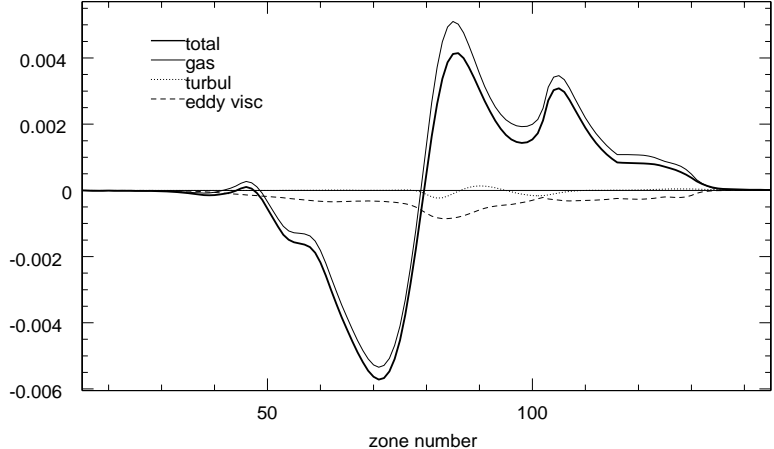,width=10.5cm}}
\caption{\small Nonlinear Work Integrand (surface to the right) showing 
separate contributions of the gas, turbulent and eddy viscosity pressures.}
 \label{nlwk}
\end{figure}

The total work (Eq.~\ref{eqwk}) that is done over a limit cycle is zero, but
again it is of interest to see how nonlinear effects change the nonlinear work
integrand which is displayed in Fig.~\ref{nlwk}.  Our nonlinear work integrand
is arbitrarily normalized by twice the nonlinear pulsational kinetic energy.
The figure shows the separate contributions of $p_g$, $p_t$ and $p_\nu$.  Here
there is, in addition, a pseudo-viscous pressure whose contribution is very small
compared to that of the other pressures and is not shown here.

In comparison to the linear work integrand, most noticeable are (a) the
broadening of the driving region because the (non Lagrangean) ionization fronts
sweep through the envelope during the pulsation; this is already known from
purely radiative models (\eg Figs.~4 and 7 in Buchler 1990) and (b) the
greatly enhanced damping by the eddy viscosity pressure $p_\nu$.

A final comment concerning frequently made approximations.  In many early
pulsation computations convection was assumed to be 'frozen in': Convection was
included in the computation of the equilibrium model, but all convective
quantities were held constant in the calculation of the period and of the
linear growth rates.  A similar approximation is often made in stellar
evolution computations.  In YKB we have examined this approximation and found
it to be very lacking.  The perturbation of the turbulent quantities, and
concomitantly of the convective flux, has a very strong damping effect on the
pulsation. 

\begin{figure}
\psfig{figure=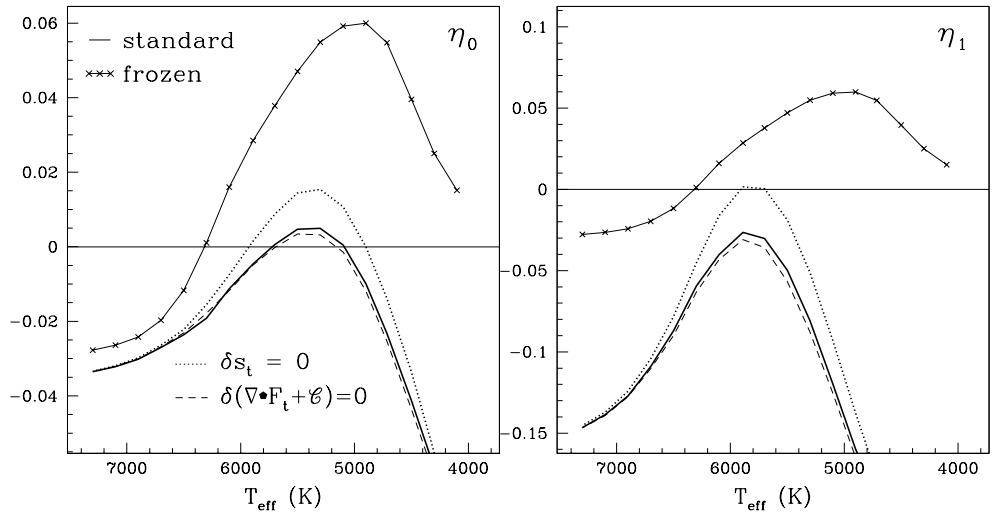,width=14.cm}
\caption{\small Effects of commonly made  approximations on position of red and
blue edges;  left: fundamental mode,  right: first overtone.} 
\label{fig_eta_approx}
\end{figure}

 In Fig.~\ref{fig_eta_approx} we summarize these results for a \Teff sequence
of models (with $M$=5\Mo, $L$=2060\Lo) for the fundamental and first overtone
modes.  The solid lines represent the exact growth rates (i.e. correct
linearization of all quantities).  The line with crosses represents the 'frozen
convection' approximation which is seen to be inadequate.  The fundamental
instability strip (the domain where the modal growth rate is positive) is
enormously broadened and shifted.  For the overtone the effect appears even
more drastic.  The mode, which is stable throughout the whole temperature
region, now becomes unstable over a very broad region.

The dotted line corresponds to the approximation of 'adiabatic' convection,
i.e. $T\delta s_t$ $\equiv $ $\delta e_t - p_t\delta v = 0$.  Physically it
corresponds to assuming that all convective time scales are very long.  This
approximation is seen to underestimate somewhat the damping effects of
convection.

Another convenient approximation is the other extreme, which is to assume that
all convective time scales are very short compared to the other time scales,
\ie from Eq.~\ref{eqete} we obtain
 \begin{equation}
  {\partial \over\partial r}\left( r^{\scriptstyle 2} F_t\right)
  - { e^\ah_t\over\Lambda} \alpha_d\th \left(e_t - S_t\right) = 0.
 \, ,
\end{equation}
 This is seen to be the best of the approximations.  It is also the simplest to
apply in evolutionary calculations in which a time independent, local mixing
length recipe is used (which would correspond here to setting in addition
$F_t=0$ or $\alpha_t=0$, \ie no diffusion of turbulent energy).

\section{Nusselt vs. Rayleigh Numbers} 

The Nusselt number is defined as Nu=$F_c/F_{cond}$, where in
our case the conductive flux is the radiative flux, and the 
Rayleigh number is Ra=$g\beta d^3TY/(\nu\chi)$. Here $g$ is the local gravity,
$d$ is the local scale height, 
$\nu$ is the kinematic viscosity and $\chi$ is the radiative conductivity.
There is general agreement that Nu should depend on Ra,  viz. Nu =
Ra\th$ ^a$,  but
there is no theoretical agreement on what the value of $a$
should be (\eg Spiegel 1971).  Some experimental results indicate that $a=0.28$
(Castaing \etal 1989), but it is not clear that they should apply to the
stellar case where the boundaries can adjust to accommodate a fixed stellar
luminosity, and where the physical quantities have strong spatial variations,
especially through the partial ionization zones.

\begin{figure}
\psfig{figure=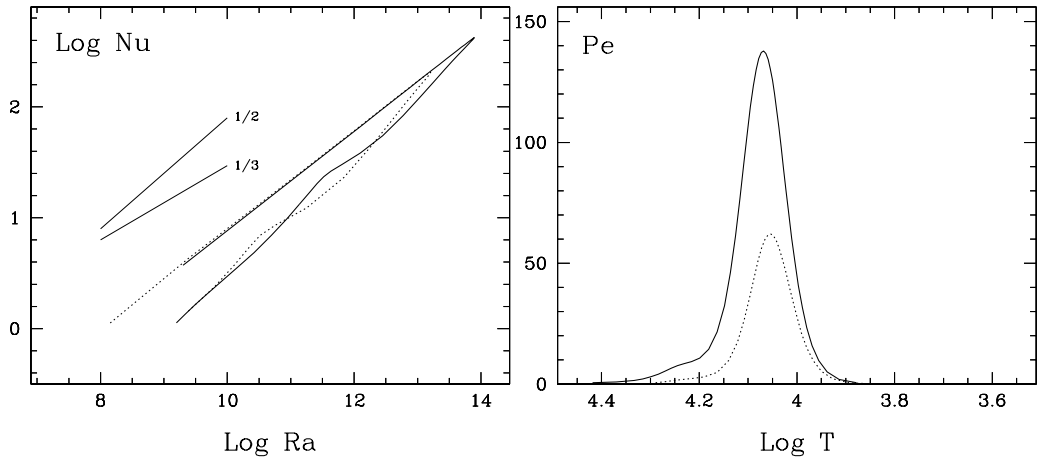,width=14.cm}
\caption{\small Left: Local Nusselt versus Rayleigh number.  Right:
P\'eclet Number in two Cepheid envelopes.} 
\label{figNuRa}
\end{figure}

In Fig.~\ref{figNuRa} we reproduce that behavior of the local Nu versus Ra
numbers throughout the convective regions of the two typical Cepheid models
$M$=5\Mo, $L$=2090\Lo, \Teff=4900 (solid line) and 5300K (dotted) of YKB.
Only the combined H--He convective regions, where Nu$>1$, are shown.  For
reference we have shown two thin lines with slopes 1/2 and 1/3, respectively.
Throughout the convective region the exponent $a$ varies between 0.45 and 0.53,
and thus agrees best with the higher theoretical value of 1/2 (Spiegel 1971).
The right hand side shows the P\'eclet number defined as the ratio of the
thermal diffusion time scale to the convective time scale.

\section{Sequence of Cepheid models}

\begin{figure}
\centerline{\psfig{figure=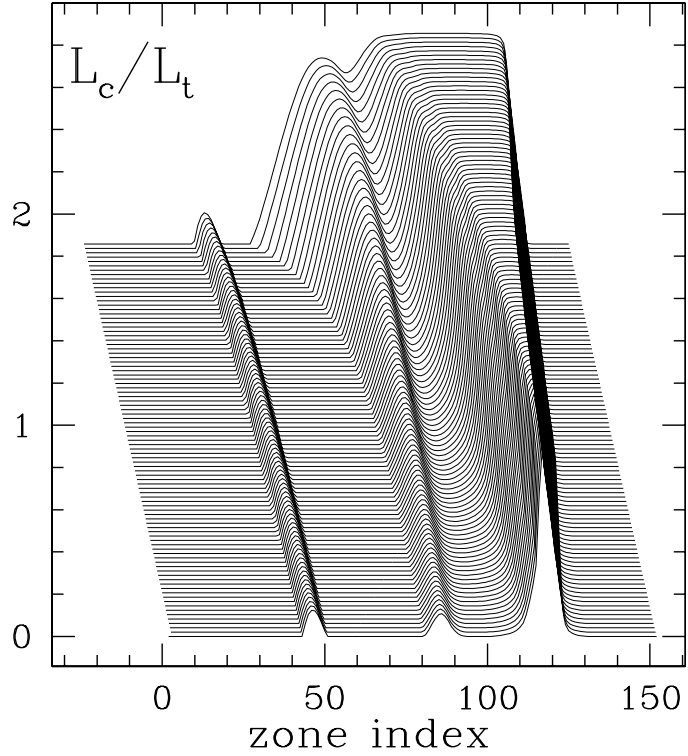,width=10cm}}
\caption{\small Profiles of the ratio of convective to total 
luminosity along a sequence; blue
edge in front (bottom)  and red edge towards the back (top).} \label{seq_fc}
\end{figure}

YKB performed some sensitivity tests of the properties of Cepheid models
obtained with the 1D turbulent convective model.  Here we just add
Fig.~\ref{seq_fc} which displays the strength of the convective luminosity as a
function of zone number (bottom scale), for a sequence of Cepheid models
starting from the blue edge in front, to the red edge in the back.  The
importance of the convective flux increases from the blue edge, where it is
relatively unimportant, to the red edge.  The H and first He ionization regions
are always merged into a single zone.  Near the blue edge the second
ionization region for He forms a separate convective region, but when we arrive
at the red edge, convection encompasses both H and He regions, and almost joins
with Fe region (left).

\section{Double-Mode (DM) Pulsations}

The numerical modelling of double mode (DM) pulsations has been a long standing
quest in which purely radiative models have failed.  In a recent paper
(Koll\'ath \etal 1998, hereafter KBBY) it was shown that with the inclusion
of turbulent convection DM pulsations appear almost naturally in Cepheid
models.  Almost concomitantly, but independently, Feuchtinger (1998) found DM
behavior in RR Lyrae pulsations which we have since also confirmed.  KBBY
described the behavior of the DM Cepheids in terms of truncated amplitude
equations (Eqs. 1 of KBBY), and they appeared to give excellent agreement with
the model that was studied.

Fig.~1 of KBBY showed the transient evolutions for a given Cepheid model and
for different initializations of the hydrocode.  The evolution toward a DM
pulsational state is clearly exhibited.  The results of the pulsational states
of a number of Cepheid models were summarized in a bifurcation diagram (Fig.~4
of KBBY).  The DM states were obtained with the regular hydrodynamics code
after lengthy time integrations with suitable initial conditions.  The single
mode pulsational states, whether stable or not, were obtained with
Stellingwerf's relaxation method (\cf Kov\'acs \& Buchler 1987), sometimes with
a lot of perseverance.

\begin{figure}
\centerline{\psfig{figure=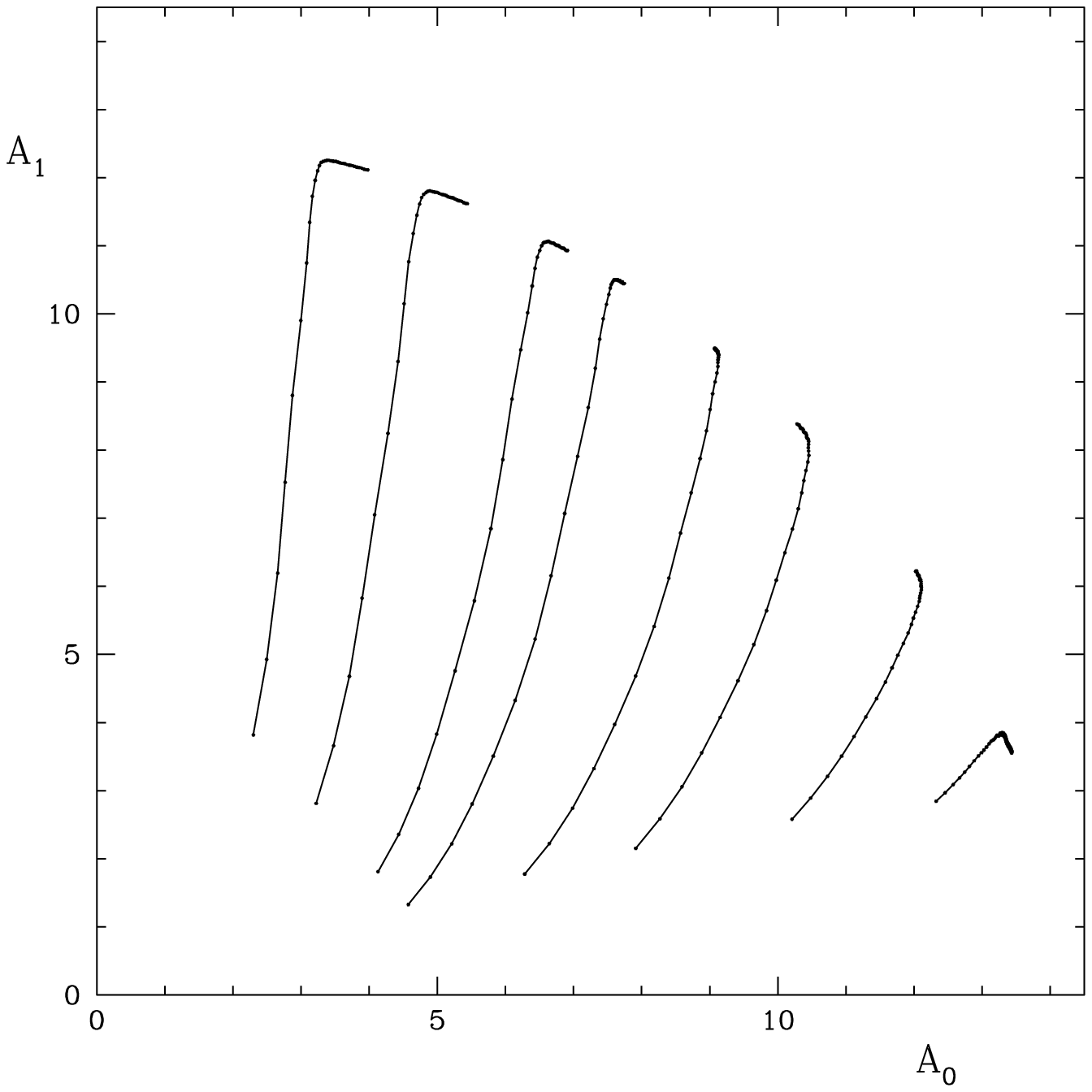,width=6.5cm}\hskip 0.5cm 
\psfig{figure=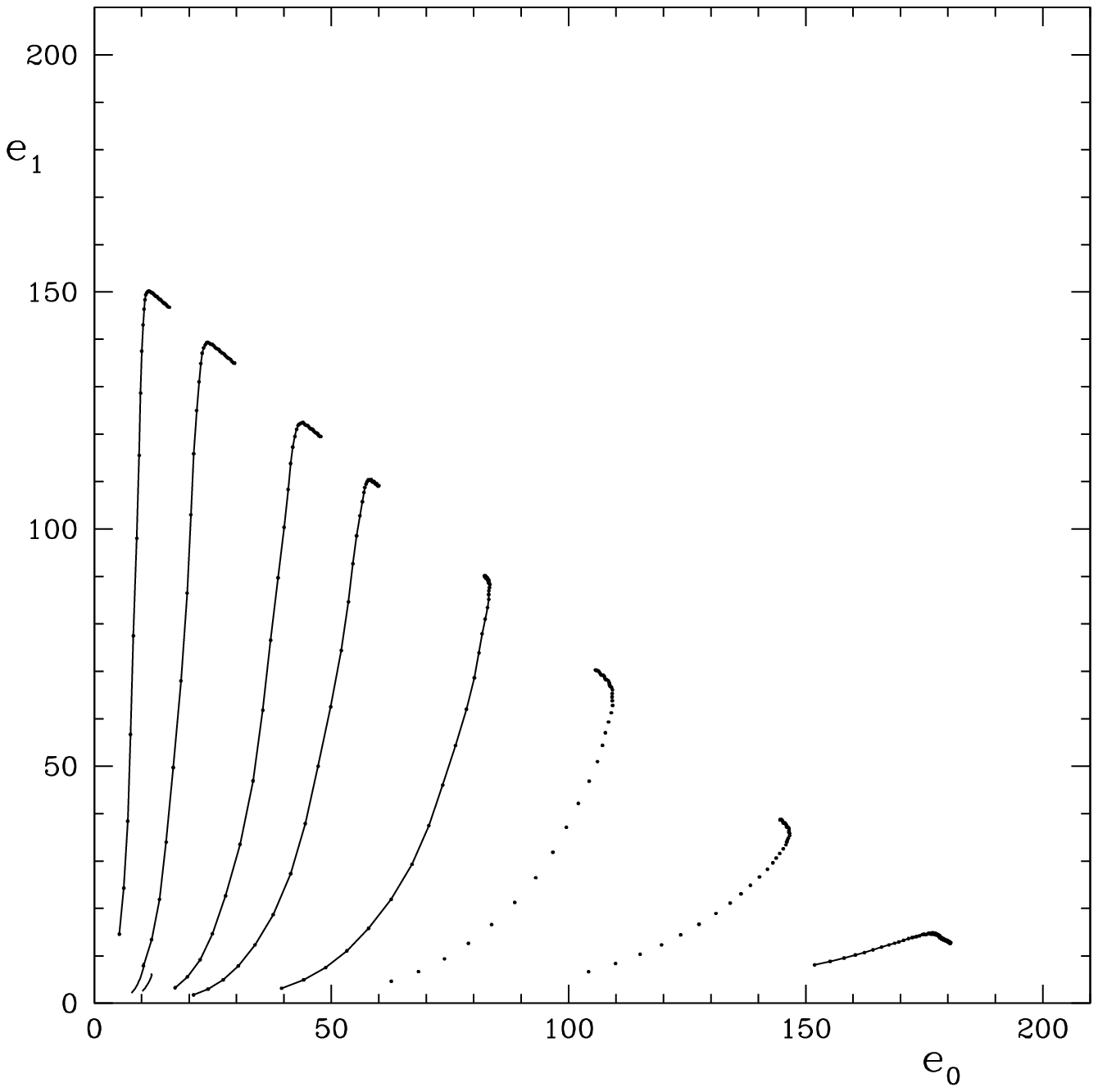,width=6.5cm}}
\vskip -6pt
\caption{\small Transient evolution  
of Cepheid model. Left (right): the F amplitude  -- 
O1 amplitude plane; right: energy plane
The trajectories correspond to various initializations.}
\label{aadm}
\end{figure}

When such models were more carefully scrutinized it became apparent that a
different transient evolution was possible (Koll\'ath \etal 1999), namely
toward the F limit cycle on the bottom right.  This situation is shown in
Fig.~\ref{aadm}.  It is clear that in addition to the stable DM there must
coexist a stable F limit cycle and a second {\sl unstable} DM.

\begin{figure}
\psfig{figure=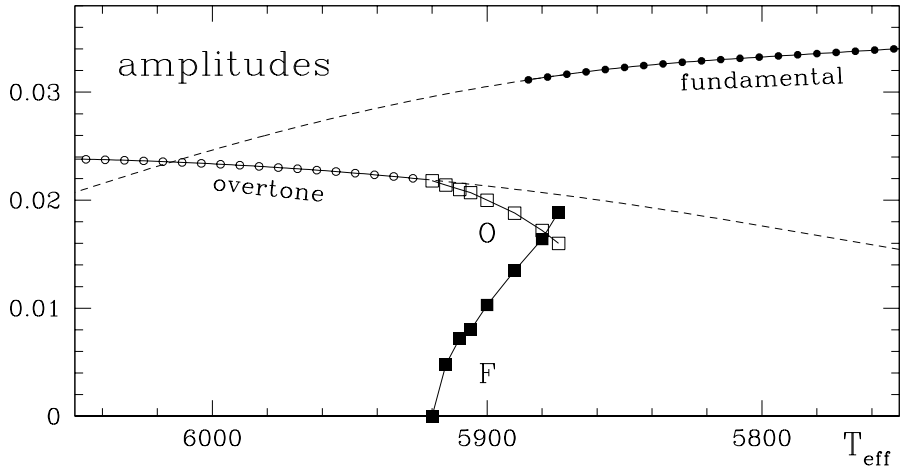,width=12cm}
\caption{\small Bifurcation diagram for the Cepheid model sequence; Solid/open
circles: F and O1  limit cycles,  Squares:  F and O1 component amplitudes of
double mode cycles.}
\label{cephbf}
\end{figure}

This new development forces us to also reconsider the bifurcation diagram
(Fig.~4 of KBBY).  We had suggested that the sharp vertical rise (drop) of the
fundamental (overtone) amplitudes was due to the presence of a pole in the
discriminant ${\cal D}$.  While such a pole is present it turns out to be too
far away (in \Teff ) to cause the observed vertical slope.  Upon closer
inspection it has been found (Koll\'ath \etal 1999) that the bifurcation
diagram is a bit more complicated than first thought.  Fig.~\ref{cephbf} is an
adaptation of the results of Koll\'ath \etal (1999).

Indeed, the region of fundamental mode pulsations extends to the left into the
region where DM  pulsations can occur.  There is thus a narrow region
of hysteresis where both F and DM pulsation can occur.  We note immediately
that this bifurcation structure, in particular the hysteresis, cannot be
accommodated with the amplitude equations of KBBY that were truncated at the
$A^3$ terms.

\begin{figure}
\centerline{\psfig{figure=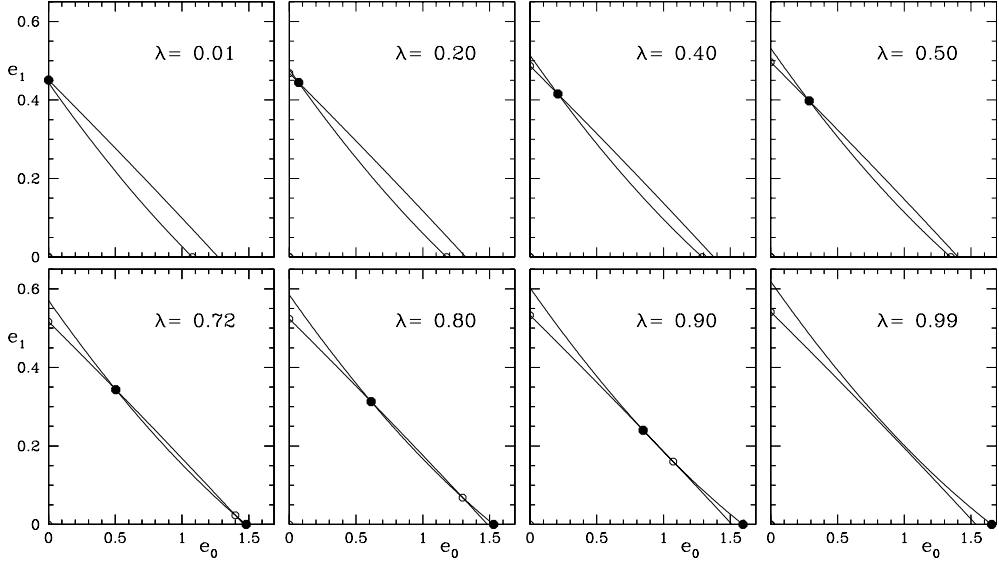,width=13.2cm}}
\vskip -5pt
\caption{The fixed point solutions as a function of the control parameter 
$\lambda$. Solid 
dots for stable, open circles for unstable fixed points.} \label{figfps}
\end{figure}

Koll\'ath \etal (1999) show that one can readily get agreement by adding the
most important next order terms in the truncation, which normal form theory shows
to be  $-r_0 A_0^5$
and $-r_1 A_1^5$.  (We disregard the additional quintic cross-coupling terms).
 \begin{eqnarray} 
 {d\th A_0\over dt}
  = \bigl (\kappa_0 -q_{00} A_0^2 -q_{01} A_1^2 \th -r_0 A_0^{\scriptstyle 4}
  \bigr )  \th\th A_0 \\ 
{d\th A_1\over dt}
  = \bigl (\kappa_1 -q_{10} A_0^2 -q_{11} A_1^2 \th -r_1 A_1^{\scriptstyle 4}
 \bigr ) \th\th A_1
 \end{eqnarray} 
 Rather than using amplitudes $A$, it is equivalent
and perhaps more convenient here to introduce the 'energies',
$e=A^{\scriptstyle 2}$, instead of the amplitudes $A$. The amplitude equations,
with the new terms added, take on the form 
\begin{eqnarray} 
{d\th e_0\over dt}
  = 2\th \bigl (\kappa_0 -q_{00} e_0 -q_{01} e_1 \th -r_0 e_0^{\scriptstyle
2}\bigr ) \th\th e_0 \\ 
{d\th e_1\over dt} = 2\th \bigl (\kappa_1 -q_{10} e_0
   -q_{11} e_1 \th -r_1 e_1^{\scriptstyle 2}\bigr ) \th\th e_1
 \end{eqnarray} 
 The loci of the fixed points are obtained by setting the RHSs of these
equations equal to zero.  Without the $r$ terms these nullclines (other than
the two coordinate axes) are simply straight lines that intersect at most once,
and when they intersect they give the DM as has been known for a long time
(Buchler \& Kov\'acs 1986).  Clearly no hysteresis is possible in this case.
On the other hand, even with small $r$ values the lines bend and multiple
intersections become possible.

This situation is depicted in Fig.~\ref{figfps}.  For the sake of illustration,
we have chosen the numerical values ($q_{00}$=2.179e-3, $q_{01}$=4.5e-3,
$q_{10}$=5.9e-3, $q_{11}$=16.e-3, $r_0$=--3.e-4, $r_1$=1.4e-3), for
simplicity keeping these values constant even though in a real sequence of
models they would vary.  The variation of the growth rates along a sequence is
more important and we assume that $\kappa_0 = \bar\kappa_0$ + 0.8e-3$\lambda$,
$\kappa_1 =$ $\bar\kappa_1$ + 1.6e-3 $\lambda$, where $\bar\kappa_0$=2.e-3,
$\bar\kappa_1$=7.5e-03.  The parameter $\lambda$ varies between 0 and 1 along
this sequence.

\begin{figure}
\psfig{figure=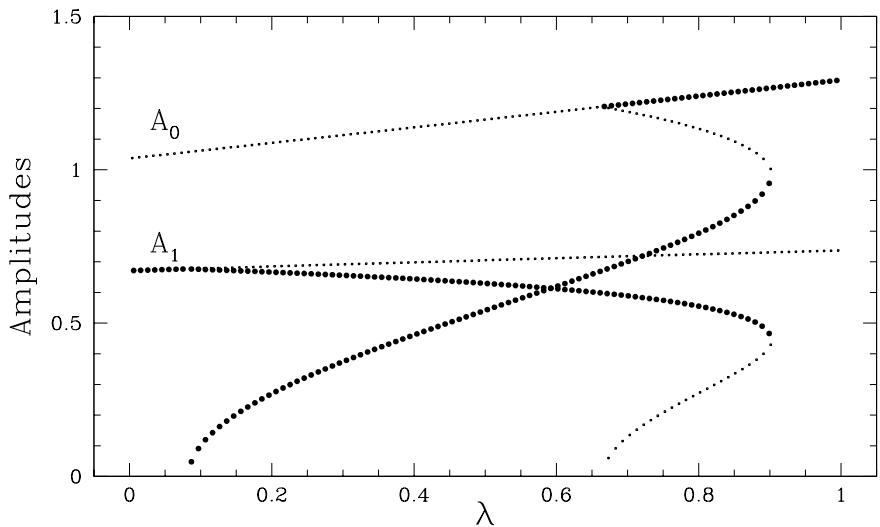,width=12cm}
\caption{Bifurcation diagram corresponding to the illustrative example.
Amplitudes of the SM and DM limit cycles.  The stable (unstable) limit cycles
are denoted by thick (thin) dots.}
\label{modbd}
\end{figure}

The corresponding bifurcation diagram is presented in Fig.~\ref{modbd}.  It is
seen to display the same general features as the actual Cepheid diagram.  In
particular, it has an single-mode O1 regime up to $\lambda\sim$~0.08, a DM
regime from $\lambda\sim$~0.08--0.90, and a coexistence between DM and F modes
from $\lambda\sim$0.67--0.90.  To the right $\lambda\approxgt$~0.90, only the F
mode LC is stable.  Note that the annihilation of the stable and unstable DMs
that occurs at $\lambda\sim$~0.90 gives rise to the vertical observed tangent.


 Note that the complexity of the bifurcation diagram is partly due to the
values we have chosen for the control parameters, \Teff\ and $\alpha_\nu$.
Our values correspond to realistic Cepheids and are not idealized for the
purpose of clarifying the evolution into single and double modes.  If we had
chosen instead to unravel the complete nature of the bifurcation, we would have
been forced to choose {\sl both} \Teff\ and $\alpha_\nu$ to correspond to the
polycritical point -- where the F, O, DM and the trivial solution coexist (this
point was previously discussed in KBBY and plotted there in Fig.~3).  Near to
this point the dynamics is given by the cubic equations (we can take this as
the definition of near).  Furthermore, the bifurcation structure is
straightforward once we know how we move through the parameter space given by
\Teff\ and $\alpha_\nu$. (This is not true if the bifurcation is subcritical,
but as yet we have not encounterd this case).  For a more general unfolding of
the bifurcation, however, this ideal picture is easily extended beyond its
reach.  So we ought to expect that some effects of this breakdown, in the form
of an increasing nonlinearity, will begin to appear.  What we seem to be
witnessing here is, in fact, the need for quintic terms as the polycritical
point becomes more distant.

\section{Conclusions}

It is perhaps remarkable that such a simple 1D recipe for turbulent convection
can give such drastic improvements over purely radiative codes.  It may
indicate that, at least for Cepheid and RR Lyrae variables, this recipe
incorporates all the physics of turbulence and convection that is essential to
model these pulsations.  It is possible that a nonlocal, time dependent
dissipation which this model equation provides is all that is needed.

We are only at the beginning of the process of calibrating the seven $\alpha$
parameters that appear in the turbulent convective description.  There are
numerous constraints that need to be satisfied, and we hope that despite the
large number of these $\alpha$'s these constraints can be satisfied.  In
particular it will be a challenge to obtain the observational properties of
both the Galactic and of the Magellanic Cloud variable stars.  Only then will
we know whether our simple 1D model is adequate.

\acknowledgments

We wish to thank the organizers for a most pleasant and fruitful meeting.  This
work has been supported by NSF (AST95--28338).

 \end{document}